\documentclass[aps,pra,showpacs,groupedaddress,twocolumn]{revtex4}
\usepackage{amsfonts}
\usepackage{amsmath}
\usepackage{mathrsfs}
\usepackage{amssymb}
\usepackage[dvips]{graphics,color}
\usepackage{graphicx}
\usepackage{dcolumn}
\usepackage{bm}
\begin{document}
\title{Laser-catalyzed spin-exchange process in a Bose-Einstein condensate}
\author{H. Jing$^{1,2}$, Y. Jiang$^1$, Weiping Zhang$^2$, and P. Meystre$^{3}$}
\affiliation{$^1$Department of Physics, Henan Normal University,
Xinxiang 453007, People's
Republic of China\\
$^2$ State Key Laboratory of Precision Spectroscopy, Department of
Physics, East China Normal University, Shanghai 200062, People's
Republic of China\\
$^{3}$B2 Institute and Department of Physics, The University of
Arizona, Tucson, Arizona 85721, USA}
\date{\today}
\begin{abstract}
We show theoretically that it is possible to optically control
collective spin-exchange processes in spinor Bose condensates
through virtual photoassociation. The interplay between optically induced
spin exchange and spin-dependent collisions provides a flexible tool for the
control of atomic spin dynamics, including enhanced or inhibited
quantum spin oscillations, the optically-induced
ferromagnetic-to-antiferromagnetic transition, and coherent
matter-wave spin conversion.
\end{abstract}
 \maketitle

Spinor Bose-Einstein condensates, consisting of atoms with internal
spin states, provide a promising bridge between atomic, molecular and optical physics (AMO), matter-wave optics, many-body physics, and
quantum information science \cite{PM}. Beside the study of their
rich ground-sate properties \cite{Ho} and spatial spin structures
\cite{Wen,Gu,sadler}, the magnetic control of quantum spin mixing
has been a central topic of investigations for atomic spin systems
\cite{Law,Plimak,Lee,Chang,Zhou,J,M,Black}. Also, by steering the
spin degrees of freedom in ultracold quantum gases, the creation of
topological skyrmions, Dirac monopoles, dark solitons, and quantum
entanglement have been investigated \cite{Leanhardt,Du,Widera}.

In parallel to these works, which concentrate largely on the role of
external magnetic fields on spinor condensates, there have also been
important developments on their magneto-optical manipulation. For
example, Zhang $et~al.$ \cite{wei} studied the spin waves induced by
light-induced dipole-dipole interactions in an atomic spin chain
trapped in a lattice potential. More recently, advances in
coherent photoassociation (PA) at ultracold temperatures \cite
{Winkler} were exploited in theoretical and experimental studies of
spin mixing and spin-dependent PA in spin-1 condensates
\cite{Cheng,Chapman}. In very recent work, Kobayashi $et~al.$
observed the spin-selective formation of spinor molecules in
ferromagnetic atoms $^{87}$Rb \cite{kobayashi}.

In this Rapid Communication, we demonstrate theoretically the
optical control of atomic spin mixing in a ferromagnetic spin-1 Bose
gas. In the proposed method the optical fields induce virtual PA
processes in the atoms, for example via a dark molecular state. We
show that this step, which can be intuitively coined a
laser-catalyzed spin-exchange process (LCSE), opens up an efficient
and well-controlled optical channel for coherent atomic spin mixing.
By tuning the the strength ratio of these two channels of
LCSE and spin-dependent collisions, three different regimes can be
identified in the laser-controlled quantum spin dynamics, i.e.,
going from the collision-dominated regime to the no-spin-mixing
regime, and to the laser-induced antiferromagnetic regime, which is
reminiscent of the prominent role of long-range dipole-dipole
interaction in a dipolar spin gas (by changing the ratio of
spin-dependent collision and dipolar interaction). As a result, and
in contrast to the collision-dominated "single-channel" case, a
wealth of important new effects arise in the spin dynamics of the
atoms.

For concreteness we concentrate on two limiting situations, the
adiabatic far off-resonant regime and the resonant case \cite{Hong}.
Our purpose here is to show that the interplay of two channels, the
spin-dependent collisions and LCSE, provides a flexible
tool for the control of the atomic spin dynamics, including enhanced
and inhibited quantum spin oscillations, laser-induced
ferromagnetic-to-antiferromagnetic (F-AF) transitions, as well as
efficient coherent spin transfer triggered even by quantum vacuum
noise. This method can be also extended to study e.g. the optical
control of domain formation and of spin textures in a spinor gas
\cite{sadler}. As such, optical LCSE-controlled spinor condensates
provide a promising new tool for the study of collective
chemically-driven quantum spin dynamics.

\begin{figure}[ht]
\includegraphics[width=90mm]{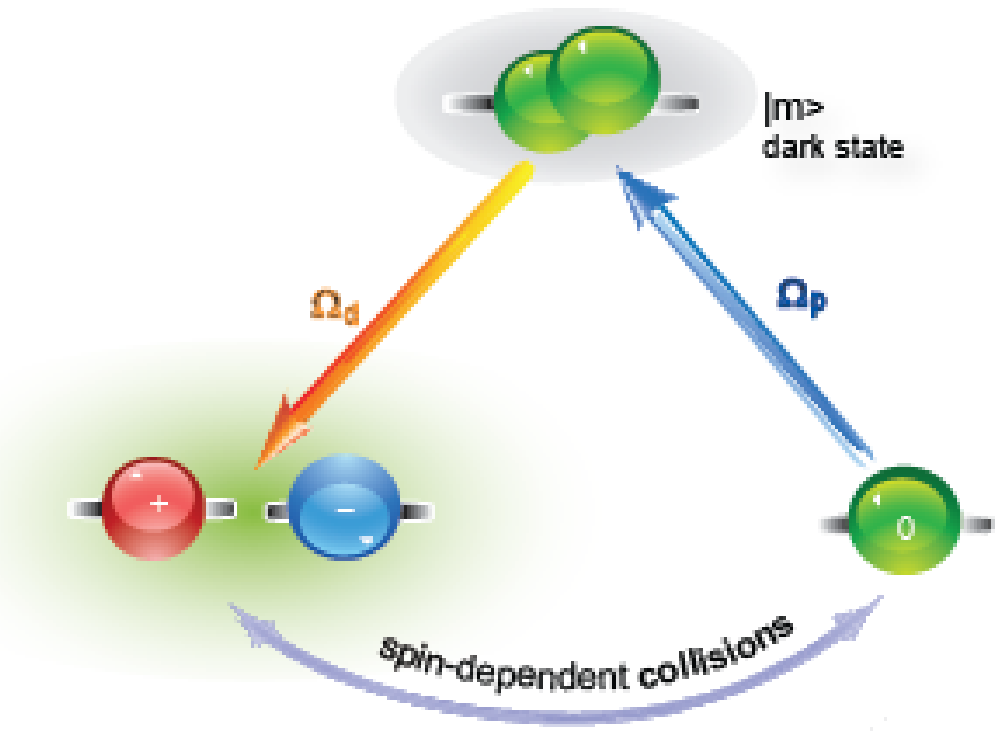}
\caption{(Color online) Schematic of coherent two-channel
spin-exchange interactions in an optically-controlled spin-1 Bose
gas. In addition to the familiar collisional channel, the optical
channel proceeds via the virtual PA of two $|m_F=0\rangle$ atoms
into an intermediate molecular state $|m\rangle$, which dissociates
into a pair of $|m_F=+\rangle$ and $|m_F=-\rangle$ atoms.}
\end{figure}

Figure~1 illustrates the process under consideration, the dynamics
of a spin-1 atomic condensate resulting from the virtual PA of two
$m_F=0$ atoms into a molecular state, followed by dissociation into
a pair of $m_F=-1$ and $m_F=+1$ atoms. Accounting in addition for
spin-dependent collisions between atoms, this system is described by
the Hamiltonian ($\hbar = 1$)
\begin{align}
H=H_{\rm coll}+H_{\rm pa},
\end{align}
where $H_{\rm coll}$ and $H_{\rm pa}$ refer to
spin-dependent collisions and the controllable light-assisted
interactions, respectively, with
\begin{eqnarray}
H_{\rm coll}&=&\int d{\bf r}\Big[\hat\psi_i^\dag
(-{\frac{\hbar^2}{2m}}\nabla^2 + V+E_i) \hat\psi_i +\frac{c_0'}{2}
\hat\psi_i^\dag\hat\psi_j^\dag\hat\psi_j\hat\psi_i\nonumber\\
&+&\frac{c_2'}{2} \hat\psi_k^\dag\hat\psi_i^\dag(F_\gamma
)_{ij}(F_\gamma)_{kl}\hat\psi_j \hat\psi_l\Big],\\
 H_{\rm pa}&=&\int
d{\bf
r}\Big[\Delta'\hat\psi^\dag_m\hat\psi_m+\Omega_p(\hat\psi^{\dag
2}_0\hat\psi_m+\hat\psi^\dag_m\hat\psi^2_0)\nonumber\\
&-&\Omega_d(\hat\psi^{\dag
}_m\hat\psi_+\hat\psi_-+\hat\psi^\dag_-\hat\psi^\dag_+\hat\psi_m)\Big].
\end{eqnarray}
Here $\hat\psi_i(\rm r)$ is the annihilation operator of the $i$-th
component atom ($i=0,\pm 1$) and $\hat\psi_m(\rm r)$ the
corresponding operator for the intermediate molecular field,
 $V$ is the trap potential, $E_i$ is the Zeeman shift and $F_{\gamma=x,y,z}$ is the spin-1 matrix \cite{Wen}. The coefficients $c'_0=4\pi\hbar^2(a_0+2a_2)/3m$ and
 $c'_2=4\pi\hbar^2(a_2-a_0)/3m$ give the strength of the spin-conserving two-body collisions, with $a_0$, $a_2$ the scattering lengths of the accessible collision channels, and $m$ is the atomic mass. The Rabi frequencies $\Omega_{p,d}$ describe the strength of the photoassociation of $m_F=0$ atoms and dissociation into $m_F=\pm1$ atoms, and $\Delta'$ is the detuning between the molecular and atomic states.

We first consider the adiabatic off-resonant regime by assuming that
$\Delta'$ is the largest parameter in the system. From
${\partial{\hat\psi}_m}/{\Delta'\partial t}\simeq 0$, we have
$\hat\psi_m\simeq\frac{\Omega_d\hat\psi_+\hat\psi_--\Omega_p\hat\psi^2_0}{\Delta'}$,
and substitute this form into the Heisenberg equations of motion
derived from Eq.~(1) \cite{maschler}. It is easily seen that the
resulting equations can also be derived from the effective
Hamiltonian
\begin{equation}
\mathcal{H}_{\rm eff}=H_{\rm coll}+\int d{\bf
r}\Big[\Omega'(\hat\psi^\dag_+\hat\psi^\dag_-\hat\psi^2_0
+\hat\psi^{\dag2}_0\hat\psi_+\hat\psi_-)+\mathcal{O}\Big],
\end{equation}
with
$$\Omega'={\Omega_p\Omega_d}/{\Delta'},~~~~\mathcal{O}=
-\frac{\Omega^{2}_p}{\Delta'}\hat\psi^{\dag2}_0\hat\psi^2_0
-\frac{\Omega^{2}_d}{\Delta'}\hat\psi^\dag_-\hat\psi^\dag_+\hat\psi_+\hat\psi_-.$$
In addition to the familiar spin-dependent collisions, spin coupling
now also results from optical LCSE, resulting in the two-channel
spin-exchange Hamiltonian
$$\mathcal{H}=\mathcal{C}'\int d{\bf
r}(\hat\psi^\dag_+(\mathbf{r},t)\hat\psi^\dag_-(\mathbf{r},t)\hat\psi_0(\mathbf{r},t)\hat\psi_0(\mathbf{r},t)+h.c.).$$
Here $\mathcal{C}'=\Omega'+c'_2$ describes the combined effect of
spin-dependent collisions and optical LCSE. Clearly, the effective
spin-coupling strength $\mathcal{C}'$ can be negative or positive
with suitable optical parameters, leading to significantly different
spin dynamics.

In the limit where the spatial degrees of freedom decouple from the
spinor dynamics, that is, when the spin healing length is larger
than the condensate size, it is possible to invoke the single-mode
approximation \cite{Chang,Black} where
$\hat\psi_i(\mathbf{r},t)\rightarrow\phi(\mathbf{r})\hat a_i(t),~
\hat\psi_m(\mathbf{r},t)\rightarrow\phi(\mathbf{r})\hat m(t)$, where
$\phi(r)$ is the spatial wave function of the condensate with $\hat
a_i$ and $\hat m$ being the atomic or molecular annihilation
operators. The rest of this paper presents results obtained in this
approximation. For convenience we also introduce the scaled
parameters $c_{0,2}=c'_{0,2}\int d{\bf r}|\phi({\bf r})|^4$,
$\Omega=\Omega'\int d{\bf r}|\phi({\bf r})|^4$,
$\Delta=\frac{\Omega_d^2}{\Delta'}\int d{\bf r}|\phi({\bf r})|^4$,
and $\tau=c_0nt$, where $n$ is the initial atomic density.

The mean-field evolution of the spin-$0$ atomic population $n_0$ is
illustrated in Fig.~2 for several values of $\mathcal C=\Omega+c_2$
and for the initial state $(n_+,~n_0,~n_-)=(0.05,~0.9,~0.05)$. The
specific example of atoms $^{87}$Rb is considered here, with
$a_0=(101.8\pm2)a_B$ and $a_2=(100.4\pm1)a_B$ \cite{kempen}, where
$a_B$ is the Bohr radius. In our calculations, we assume the optical
detuning $\Delta'=100\Omega_p$ and $\Omega_d=10\Omega_p$, which is
reasonable for the adiabatic off-resonant case. The typical initial
atomic density $n \sim10^{14}cm^{-3},$ corresponding to $c_0n \sim
9700$Hz.

Significantly different regimes are reached by varying the optical
detuning $\Delta'$: (i) In the collision-dominated regime
$|c_2|>|\Omega|$, the population of spin-$0$ atoms is always
$larger$ than its initial value $0.9$. In this perturbed regime, the
spin coupling is still ferromagnetic ($\mathcal{C}<0$); (ii) For
$\mathcal C=0$, i.e. $\Omega=-c_2$ (for $\Delta'>0$) the two
channels (spin-dependent collisions and LCSE) interfere
destructively, leading to frozen or inhibited spin mixing; (iii) The
sign of $\mathcal C$ can be reversed by tuning the laser fields,
resulting in an effective antiferromagnetic regime $\mathcal C>0$.
This is reminiscent of the F-AF transition induced by long-range
dipolar interactions in a spinor gas \cite{Zhou}. In this reversed
regime, the atomic spin oscillations ($|\Omega|>|c_2|$) turn out to
be $below$ the initial value $0.9$.

The mean-field atomic spinor dynamics is well described by a
nonrigid pendulum model \cite{Wen}. By expressing the $c$-number
amplitudes $a_i$ in terms of real amplitudes and phases, i.e.
$a_{\pm,0}=\sqrt{n_{\pm,0}}e^{-i\theta_{\pm,0}}$, the energy
functional of the optically-controlled spinor system can be derived
as
\begin{eqnarray}
{\cal
E}&=&q(1-n_0)+\mathcal C\sqrt{(1-n_0)^2-m^2}cos\theta\nonumber\\
&+&c_2n_0(1-n_0)
+\frac{\Delta}{4}n_0(2-n_0)-\frac{\Omega^2}{\Delta}n_0^2
\end{eqnarray}
where $q$ denotes the quadratic Zeeman effect,
$\theta=\theta_++\theta_--2\theta_0$ is the relative phase of the spin components and $m=n_+-n_-$ is the atomic magnetization. The last two terms in Eq.~(5) account for the optical energy shift.

The effective spin-coupling parameter $\mathcal C$ has an important
impact on the properties of the system, as already mentioned.
Figure~3 plots equal-energy contours in the phase space
$(\theta,n_0)$, for several values of $\mathcal C$ for $m=0$ and
$q=0.01$, corresponding to the fixed magnetic field about $460mG$ in
our specific example, which is larger than the resonance magnetic
field $B_{\rm res}\sim$ 330 mG, see \cite{res}. We have also
carried out numerical simulations for values of $q$ corresponding to
$B<B_{\rm res}$ and found similar results for appropriate values of
$\mathcal C$. For our parameters only open trajectories exist in the
absence of laser fields ($\mathcal C=c_2$) as well as for perturbed
case $\mathcal C=0.5 c_2$. In contrast, the "reversed" cases
$\mathcal C=-0.5 c_2$ and $\mathcal C=- c_2$ are characterized by
the coexistence of closed and open trajectories.

\begin{figure}[ht]
\includegraphics[width=0.45\textwidth]{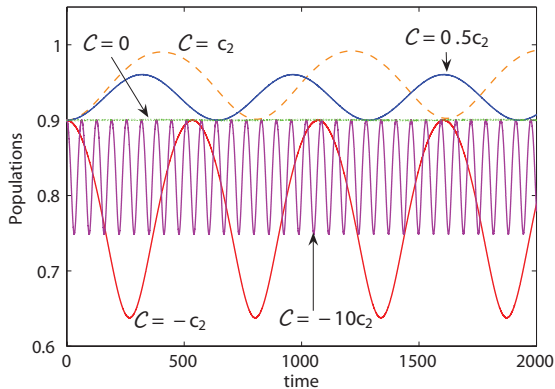}
\caption{(color online) Population of spin-zero atoms with (solid
lines) or without (dashed lines) laser fields. In all cases the
atomic density is $2\times10^{14}$ cm$^{-3}$ \cite{YL}, and the time
is scaled to $\tau=c_0nt$, corresponding to the time unit $0.1ms$.}
\end{figure}

We remark that Eq.~(5) also permits to study the instabilities and
domain formation in spin-1 atoms \cite{Wen,Gu}. We actually have
done this and found that the LCSE-dominated case is again different
from the collision-dominated case \cite{later}.

\begin{figure}[ht]
\includegraphics[width=85mm]{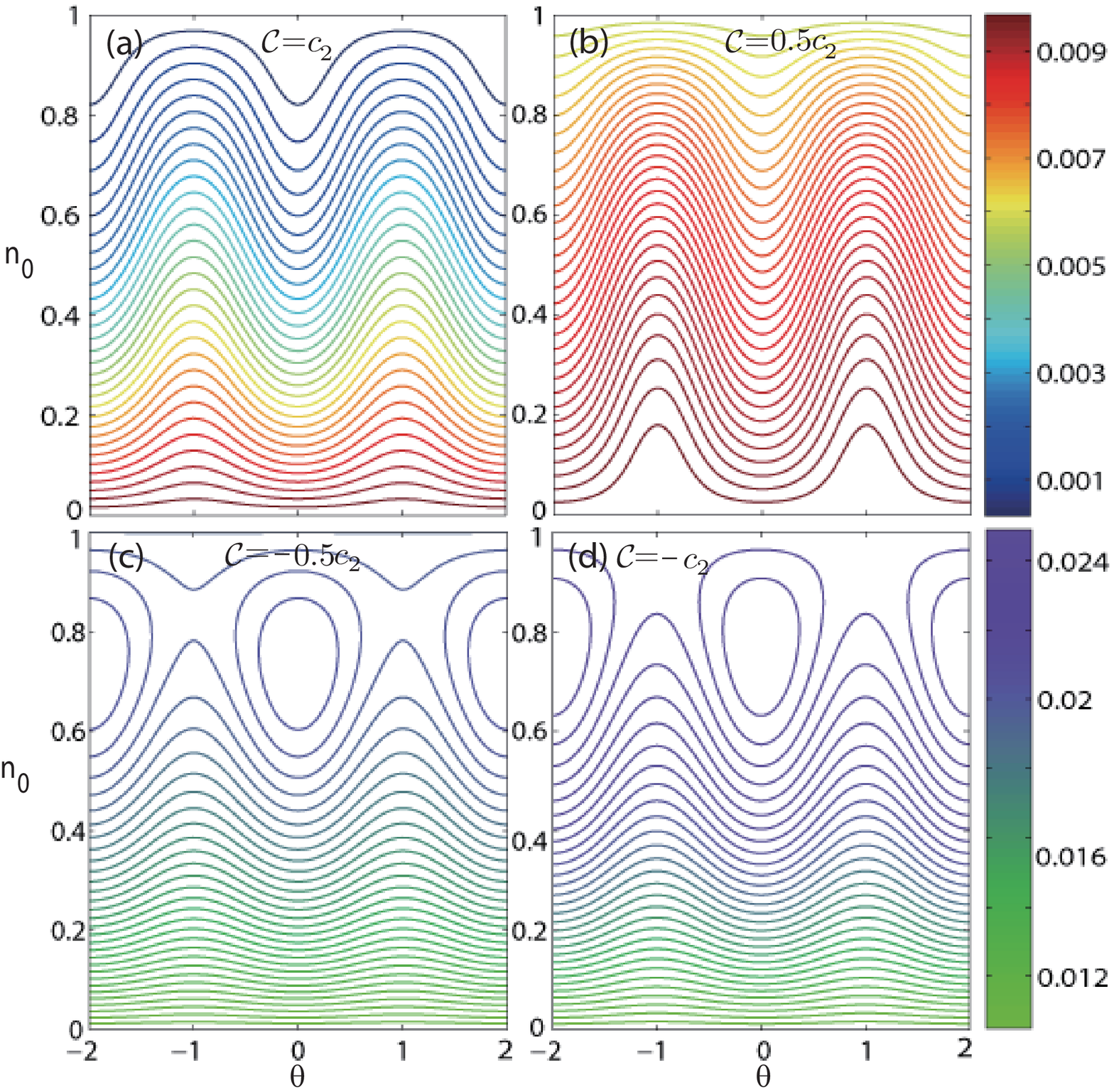}
\caption{(color online) Equal energy contours of $\cal E$ without
(a) ($\mathcal C=c_2$) and with (b) ($\mathcal C=0.5c_2$), (c)
($\mathcal C=-0.5c_2$, and (d) ($\mathcal C=-c_2$))
optically-controlled SER for $q=0.01$ and $m=0$.}
\end{figure}

We now turn to the resonant situation, where one can exploit the
existence of an atom-molecule dark state to prevent the build-up of
a significant molecular population throughout the process
\cite{Winkler}. This two-photon resonant control of atomic spinor
dynamics represents a promising new step in developing the field of
all-optical manipulation of matter-wave spins. In a mean-field
approach where $\hat{\psi_{{ i}}}\rightarrow\sqrt{n}\phi_{{ i}}$
\cite{Cheng} we find
\begin{align}
\frac{d\phi_+}{d\tau}=&-i[c_2(|\phi_+|^2+|\phi_0|^2-|\phi_-|^2)]\phi_+\nonumber\\
&-ic_2\phi_0^2\phi^*_-+i\Omega'_d\phi_m\phi^*_--i(\Theta+\delta)\phi_+,\nonumber\\
\frac{d\phi_0}{d\tau}=&-i[c_2(|\phi_+|^2+|\phi_-|^2)]\phi_0-2ic_2\phi_+\phi_-\phi^*_0\nonumber\\
&-2i\Omega'_p\phi_m\phi^*_0\nonumber\\
\frac{d\phi_-}{d\tau}=&-i[c_2(|\phi_-|^2+|\phi_0|^2-|\phi_+|^2)]\phi_-,\nonumber\\
&+i\Omega'_d\phi_m\phi^*_+,\nonumber\\
\frac{d
\phi_m}{d\tau}=&i\Omega'_d\phi_+\phi_--i\Omega'_p\phi^2_0-(i\delta+\gamma)\phi_m,
\end{align}
where $\Omega'_p=\Omega_p/c_0\sqrt{n}$,
$\Omega'_d=\Omega_d/c_0\sqrt{n}$, $\Theta=\Delta'/c_0n$,
$\delta=\delta'/c_0n$, $\delta$ is the frequency difference between
the molecular state and the atomic hyperfine states, and the
phenomenological decay rate $\gamma$ accounts for the loss of
intermediate molecules. Following the method of Ref.~\cite{Hong}, we
find that Eqs.~(6) admit a steady-state coherent population trapping
(CPT) solution ($n_m=0$),
$$n_{\pm,s}=\frac{1}{2+\Omega'_d/\Omega'_p},~~
n_{0,s}=\frac{1}{1+2\Omega'_p/\Omega'_d},$$ under the generalized
two-photon resonance condition
$$\Theta(t)=-\delta+c_2\left[2(n_{+,s}+n_{-,s})+2\sqrt{n_{+,s}n_{0,s}}-4n_{0,s}\right].
$$
This time-dependent resonance condition is determined by the CPT
steady-state values of the atomic density, which thereby can be
tuned by choosing suitable pumping and (time-dependent) dumping
laser fields.

Figure~4 shows the populations of the atomic sublevels for
$\Omega'_p=1$ (corresponding to 9700 Hz) and
$$\Omega'_d(t)=\Omega'_{d,0}{\rm sech}(t/t_0),$$ with
$\Omega'_{d,0}=40$, $t_0=20$, the other parameters being as in
Fig.~2. In contrast to the off-resonant spin oscillations, we have
now a full transfer of population from the initial state, say
$|m_F=0\rangle$ to a final coherent superposition of $|m_F = \pm
1\rangle$ hyperfine states. We have carried out numerical
simulations for a large set of initial $seeds$ of spin-$\pm$ atoms
and found that the stable spin conversion is always possible for
e.g. $\delta=3$. The departure of the spin-$\pm$ populations from
the ideal CPT value is due to the fact that only an approximate
adiabatic condition exists for the CPT state \cite{Hong}.

\begin{figure}[ht]
\includegraphics[width=0.5\textwidth]{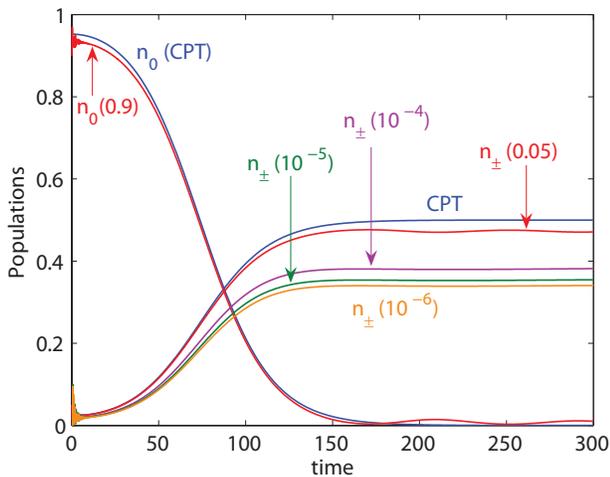}
\caption{(color online) Atomic populations in the resonant regime
for $^{87}$Rb, with detuning $\delta=3$ and $\gamma=1$. To
illustrate the role of small seeds of atoms $n_\pm$, several values
of the initial populations of the corresponding spin states have been
considered, as denoted in the brackets. The line labeled "CPT" shows
the ideal population of the target states $|m_F = \pm 1\rangle$.}
\end{figure}

Following the preparation of all atoms in the spin-$0$ state, Klempt
$et.~al$ \cite{klempt} have recently studied the magnetic field
dependence of the fraction of atoms transferred to the spin-$\pm$
states. This transfer process is of quantum nature as the mean-field
equations break down for an initial vacuum of spin-$\pm$ atoms. Our
system also permits to study quantum spin transfer with
quantum-noise-triggered $seed$. Following the strategy familiar from
quantum optics, see e.g. \cite{PM,Moore,search}, we separate the
problem into an initial stage dominated by quantum noise followed by
a classical stage that arises once the target state has acquired a
macroscopic population \cite{Li}. We have performed an analysis of
the present system along these lines, and it confirms that its
evolution from quantum noise is also efficient, with dynamics
similar to those with the classical $seed$ $n_\pm=10^{-5}$ in Fig.~5
\cite{Li,later}.

In conclusion, we have demonstrated theoretically that the interplay
of LCSE and spin-dependent collisions leads to a
variety of remarkable effects such as inhibited spin oscillations,
laser-induced F-AF transition, and quantum spin transfer between
different components. The optical tuning or even elimination of
spin-exchange interactions may be used to e.g. probe the weak signal
of dipolar interactions which is generally far smaller than the
spin-dependent collisions [10]. The optical control of atomic spin
interactions is somewhat reminiscent of the role of magnetic
Feshbach resonances in the study of ultracold gases. This work hints
at promising possibilities to carry out an all-optical control of
atomic spins \cite{spin}. When compared to the magnetic control of
spinor atoms [1-16], LCSE offers an exciting new
route to the study of atomic spin coupling independent of the
collisions. Ultimately, magneto-optical methods will likely combine
the best of both optical and purely magnetic approaches. Future work
will study the magneto-optical control of spatial spin structures
and will explore possibilities of spin-dependent ultracold chemistry
or atom-molecule hybrid spin mixing.

P.M. is supported by the U.S. Office of Naval Research, by the U.S.
National Science Foundation, and by the U.S. Army Research Office.
W.Z. is supported by the NSFC and by the 973 project. H.J. is
supported by the NSFC.


\begin{thebibliography}{99}
\bibitem{later} The details of the results will be published elsewhere.
\end{thebibliography}
\end{document}